# Joint Power and Mobility Control

Yun Hou
*Department of Computer Science*
*The Hang Seng University of Hong Kong*
Hong Kong SAR, China
aileenhou@hsu.edu.hk

Yening Zhang
*Department of Computer Science*
*The Hang Seng University of Hong Kong*
Hong Kong SAR, China
p223120@hsu.edu.hk

*Abstract—*.

## I. Introduction

Cellular Vehicle-to-Everything (V2X) is being commercialized; however, there are still challenges that need to be addressed, such as vehicular network coverage, interference, and vehicle mobility. In V2X, there are in-coverage scenarios and out-of-coverage scenarios. In the in-coverage scenario, vehicles are connected to the base station in the control plane, allowing for collision-free scheduling and high-quality service. For communication scenarios without network coverage, due to the nature of autonomous scheduling, the quality of service may be degraded under the conditions of high-density network scenarios. This is expected for future vehicular networks, especially Unmanned Aerial Vehicle (UAV) networks [1] and autonomous driving networks. In high-density network scenarios, collided message broadcasting or resource sharing may lead to significant interference, thus affecting the reception rates. Furthermore, in V2X communication scenarios, the vehicles move at high speeds, resulting in severe Doppler effects and a corresponding decrease in communication quality [2]. It is well known that network connectivity is a critical issue for future autonomous driving scenarios, as autonomous vehicles want to disseminate their status and warning messages to as many nearby vehicles as possible. Also, because of the high mobility, UAV networks are prone to possess highly dynamic, sparse, and intermittent network topologies [3]. As a result, the multi-UAV coordination or swarm operation places high demands on network connectivity.

Packet Reception Rate (PRR) is the aggregated link success probability across all receivers for one broadcast packet. Aggregated PRR over all the transmitted packets in the network, referred to as the *network PRR* hereafter, represents the level of connectivity in the network, such that the higher the network PRR is, the transmitted packets can be decoded by more vehicles in the network. However, high vehicle densities will lead to concurrent transmissions with severe interference due to the reuse of the limited radio resources between vehicles, which will degrade the PRR and hence the connectivity of the vehicular network. Therefore, in this paper, we aim to improve the network connectivity, i.e., the network PRR, via vehicle mobility (i.e., trajectory) adjustments.

The remainder of this paper is organized as follows. In Section II, we present the system model and highlight the challenges of distributed mobility control for autonomous V2X networks. In Section III, we formulate the multi-node Network Utility Maximization (NUM) problem and investigate its optimality. We devise a distributed updating framework towards the optimum in Section IV. Numerical results are shown in Section V, and Section VI concludes the paper. Our contributions are summarized in Table 1.

## II. The link reception rate model

In one subframe, a group of vehicles in the vehicular network are transmitting while others are receiving due to the half-duplexing operation. The vehicles are connected using the direction communication protocols, such as C-V2X mode 4. Therefore, they form a mesh, peer-to-peer wireless network.

The transmitting opportunity of each vehicle is self-scheduled through the Semi-Persistent Scheduling (SPS) framework [10]. All vehicles are conducting the SPS lead to a semi-static transmitter and receiver composition in the network. That is, who will be transmitting and receiving in the future can be predicted to some extent.

Our goal is to find a mechanism that allows all the transmitting vehicles to adjust their trajectory and transmission power in a coordinated and distributed manner so that the network PRR for all broadcast packets, i.e., transmitted by multiple concurrent transmitters towards all receivers in the network, is maximized.

Assume that the reception success probability of a link is proportional to the SINR of this link, while the average signal power and interference power are inversely proportional to the square or higher order of the distance associated with this link. Therefore, adjusting the distances between the transmitter and receivers is beneficial to optimize the link PRR. Due to the broadcast nature of wireless communication in the V2X context, it is not optimal to blindly increase the power or reduce the distance to a particular receiver, as this may affect other receivers. Therefore, an adjustment of distance needs to be derived from a network-wide perspective. The challenges and our solutions and assumptions in this study are summarized as follows.

In terms of the link reception of data packets, we model the packet reception rate of a link as a function of the instant link SINR. Due to the physical characteristics of the reception probability such that with extremely high and low SINR, the packet reception rate saturates towards 1 or 0, we propose to approximate the packet reception rate using a sigmoid function.

Therefore, we formulate the link reception probability as follows.in (1), $\alpha$ and $\beta$ are key parameters in shaping the sigmoid curve. Their values are to be empirically found through approximation. While, $\gamma_{i,j}$ represents the signal to noise ratios from node $j$ to node $i$ in the dB scale. The fitting of the sigmoid function's alpha and beta is presented as follows, with best alpha and beta equals (0.05, 0.525) providing an squared error of as low as 0.0039.

$$P_{i,j} = \frac{1}{1+\alpha \cdot e^{-\beta \gamma_{i,j}}}. \quad (1)$$

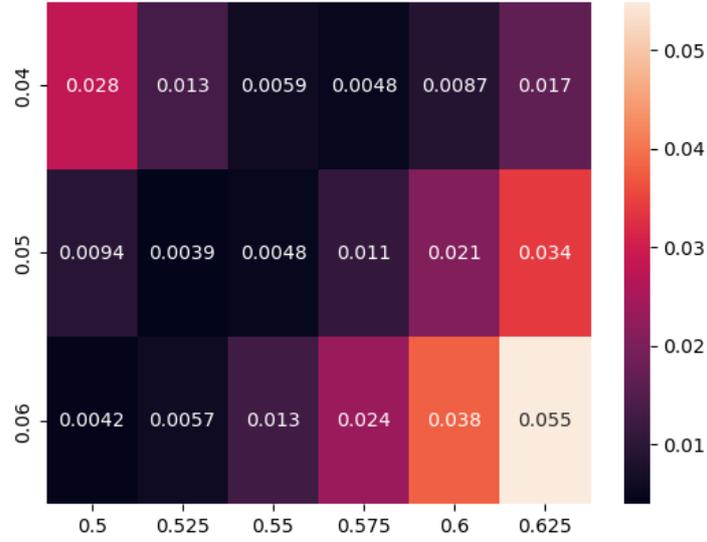

Figure 1. Fitting errors

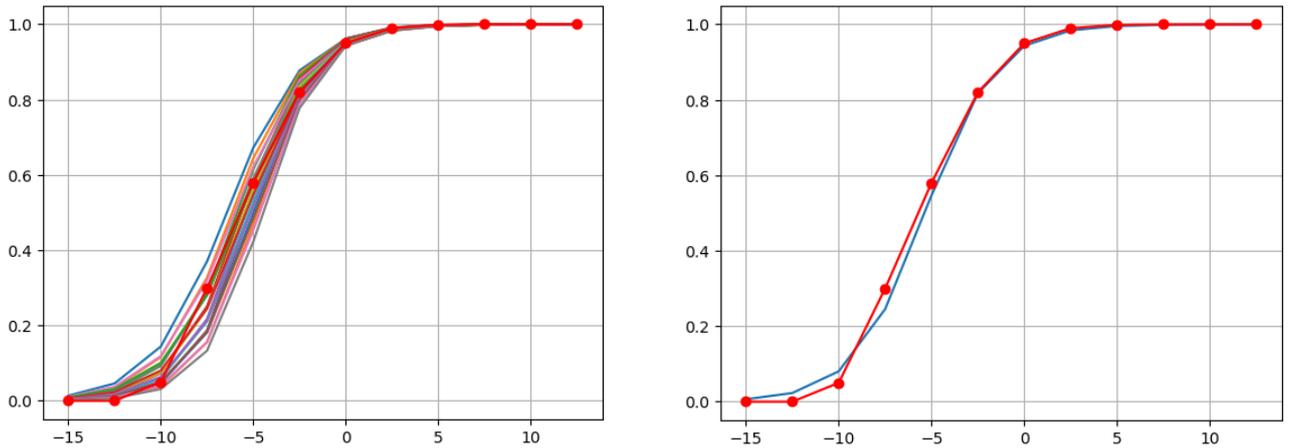

Figure 2. Fitting results

In the follwoing, we will show that the link reception model can be transformed into a function of received powers, i.e., received signal power and received interference powers. The transformation offers additional simplicity in the subsequent NUM formulation.

$$P_{i,j} = \frac{1}{1+\alpha \cdot e^{-10\beta \log_{10} \frac{g_{ij}P_{i,j}}{\sum_{k \neq j} g_{ik}P_{i,k}+N_i}}}, \quad (2)$$

$$e^{-10\beta \log_{10} \frac{g_{ij}P_{i,j}}{\sum_{k\neq j} g_{ik}P_{i,k}+N_i}}$$

$$= e^{-\frac{10}{\log_e 10}\beta \log_e \frac{g_{ij}P_{i,j}}{\sum_{k\neq j} g_{ik}P_{i,k}+N_i}}$$

$$= e^{-4.34\beta \log_e \frac{g_{ij}P_{i,j}}{\sum_{k\neq j} g_{ik}P_{i,k}+N_i}} \quad (3)$$

$$= e^{-4.34\beta \log_e \frac{g_{ij}P_{i,j}}{\sum_{k\neq j} g_{ik}P_{i,k}+N_i}}$$

$$= e^{\log_e \left(\frac{g_{ij}P_{i,j}}{\sum_{k\neq j} g_{ik}P_{i,k}+N_i}\right)^{-\beta'}}, \beta' = 4.34\beta$$

$$= \left(\frac{g_{ij}P_{i,j}}{\sum_{k\neq j} g_{ik}P_{i,k} + N_i}\right)^{-\beta'} \quad (4)$$

$$= \frac{(g_{ij}P_{i,j})^{-\beta'}}{\left(\sum_{k\neq j} g_{ik}P_{i,k}\right)^{-\beta'}}, \text{ when inteference limited.}$$

Substitute (4) back to (2), we have:

$$P_{i,j} = \frac{1}{1+\alpha \cdot \frac{(g_{ij}P_{i,j})^{-\beta'}}{\left(\sum_{k\neq j} g_{ik}P_{i,k}\right)^{-\beta'}}}, \quad (5)$$

Let $S_{i,j} = g_{ij}P_{i,j}$ denote the received signal power and $I_{i,j} = \sum_{k\neq j} g_{ij}P_{i,j}$ denote the interference power, and we then have:

$$P_{i,j} = \frac{\alpha(S_{i,j})^{-\beta'}}{\alpha(S_{i,j})^{-\beta'}+(I_{i,j})^{-\beta'}}, \quad (6)$$

It is observed that the link reception probability can be modelled as a ratio between the scaled signal power and the summation of the scaled signal power and all interference powers. This offers simplicity in our subsequence NUM formation.

### III. PROBLEM FORMULATION AND ITS CONCAVITY

We present the connectivity of a C-V2X network using the network-level aggregated PRR. Assume the utility associated with a transmitter-receiver link from node $i$ to node $j$ is proportional to the success probability of the reception $P_{i,j}$. Then, the network-level **multi-node** aggregated utility is as follows

$$U = \sum_{i\in T_x} \sum_{j\in R_x} \log(P_{i,j}), \quad (2)$$

where $T_x$ and $R_x$ represent the set of all transmitting and receiving nodes in the network, respectively.

The link success probability $P_{i,j}$ is a function of two distances according to the approximation model in (1).

The inequality (4) is added to the maximization problem to reflect the realistic constraint such that the three distances must belong to the same triangle. Equation (5) is omitted as it will not affect the second-order partial derivatives and hence not change the concavity. Thus, the optimization over $d$ is formulated as

$$\max_{d_{i,j}, i\in T_x, j\in R_x} U = \sum_{i\in T_x} \sum_{j\in R_x} \log(P_{i,j}) \quad (6)$$

To take account of the fairness among various links log (.) is applied to the reception probabilities of each link to reduce the gain in the distance when $P_{i,j}$ is already high.

A special case with number of Tx node equal to 2 can be further simplified as

$$U = \sum_i \left( \log \frac{\alpha(S_{i,1})^{-\beta'}}{\alpha(S_{i,1})^{-\beta'} + (I_{i,2})^{-\beta'}} + \log \frac{\alpha(S_{i,2})^{-\beta'}}{\alpha(S_{i,2})^{-\beta'} + (I_{i,1})^{-\beta'}} \right). \quad (6)$$

While, in this case, $S_{i,1} = I_{i,1}$ and $S_{i,2} = I_{i,2}$.

$$U = \sum_i \left( \log \frac{\alpha(S_{i,1})^{-\beta'}}{\alpha(S_{i,1})^{-\beta'} + (S_{i,1})^{-\beta'}} + \log \frac{\alpha(S_{i,2})^{-\beta'}}{\alpha(S_{i,2})^{-\beta'} + (S_{i,2})^{-\beta'}} \right). \quad (6)$$

This is transformed into a resource allocation problem for each receiver i. At each receiver i, the allocation problem is now between the Signal power from transmitter 1 and 2. It can be easily proved that the optimality exists and only exsits when $S_{i,1} = S_{i,2}$. Which means, the $g_{i1}P_{i,1} = g_{i2}P_{i,2}$ for all receiver i.

### IV. NUMERICAL AND EXPERIMENT RESULTS

In this section, we perform both Matlab simulations and on-site experiments to visualize the optimality of the NUM for two-transmitter networks.

a. Simulation results

It is verified that only when S1 equals S2, they network utility achieves the maximum.

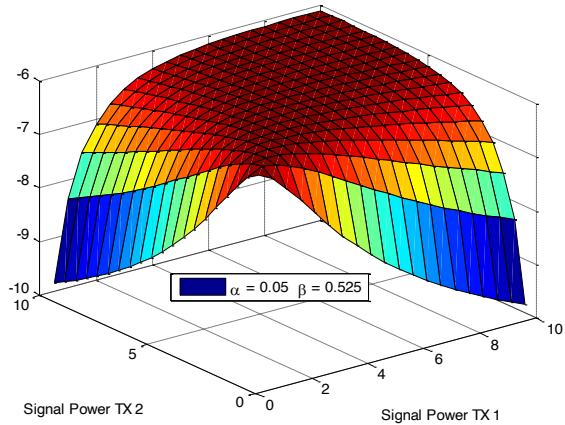

Figure 3. The average gain of 100 tests.

b. Experiment results

We use real devices to verify the 2TX network utility maximization theorem. The experiment setup consisted of three main components: robotic vehicles, positioning devices, and wireless communication modules. We conducted the experiments in the university of hall of HSUHK on 10 July, 2024. For the robotic vehicles, Yahboom robots were used, which are designed for educational and research purposes. To achieve accurate positioning, Haofang Technology UWB positioning sets were employed, including base stations and client tags, which can be easily integrated with ROS-based cars through the ROS interface. For wireless communication, ESP32 modules configured in mesh mode were utilized to enable reliable peer-to-peer connectivity among the vehicles.

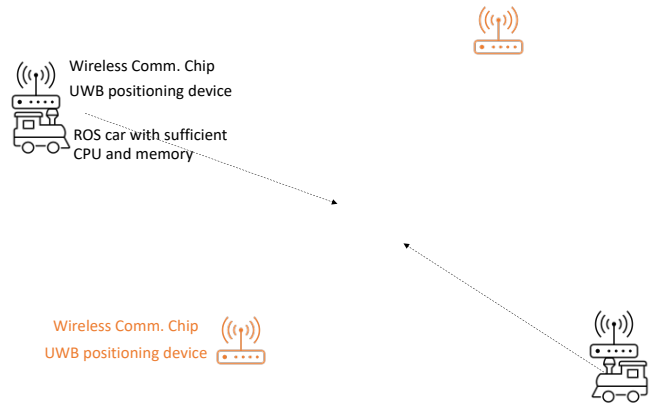

Figure 4. Setting of the experiments

The 2Tx-2RX scenario. The 2 TX are controlled by the optimization framework to update their positions (to move) to maximize the overall utility provided by the four links of (tx1, rx1), (tx1, rx2), (tx2, rx1), and (tx2, rx2).

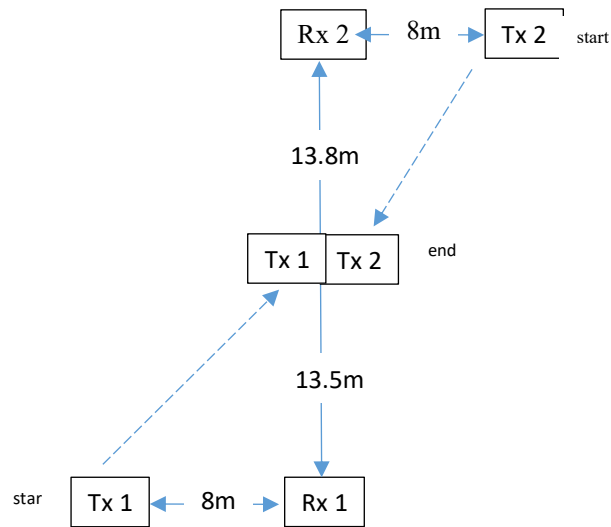

Figure 5. Trajectories of TX vechiles.

A snapshot of the robot vehicle movements as recorded by the UWB kit visualization application is shown below. On the left of the screen, we show the real-time tracked node positions. Nodes annotated as 0,1, and 2 are base station anchor points in the positioning system, while nodes 0 and 2 are also playing as the stationary receivers. On the receiver board, we use ESP chip to count and validate instant packet reception and compute averaged packet reception rate. The average is computed using smoothing average with a window length of 2 seconds. The real-time calculated packet reception rate are then displayed on the LED as shown on the right side of Fig. 6.

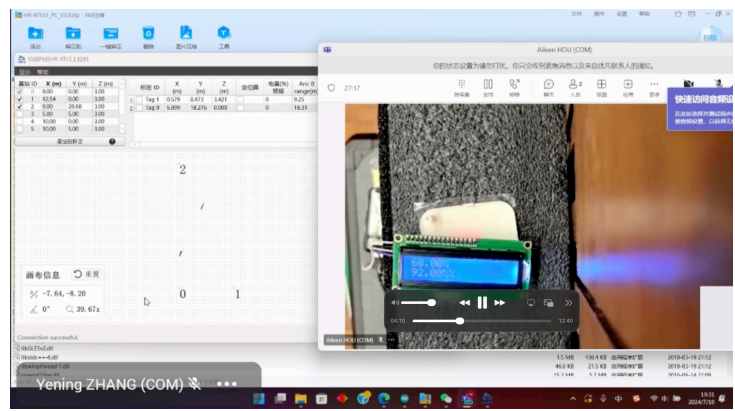

Figure 6. A snapshot of the experiment.

Finally, the packet reception rates along the moving trajectories are collected. Table 2 shows a comparison in terms of link reception rate between the start and end positions. It is verified that the symmetric position and power allocation provides the highest link reception rate in interference limited scenarios.

Table 2. Comparison of link reception rate

|  | **Start position** | **End position** |
|---|---|---|
| **Rx2, Tx2** | 0.91 | 0.88 |
| **Rx2, Tx1** | 0.76 | 0.88 |
| **Rx1, Tx2** | 0.68 | 0.90 |
| **Rx1, Tx1** | 0.93 | 0.89 |
| **Network level PRR** | 0.82 | 0.88 |

## V. Conclusion

This study addressed the challenge of improving network connectivity in autonomous V2X networks by jointly optimizing transmission power and vehicle mobility. We proposed a link reception model based on a sigmoid approximation of SINR and transformed it into a power-based formulation for simplicity in optimization. Building on this, we formulated a multi-node Network Utility Maximization (NUM) problem and demonstrated its concavity, enabling distributed trajectory and power adjustments. Both simulation and real-world experiments validated the theoretical findings, showing that symmetric positioning and balanced power allocation significantly enhance packet reception rates under interference-limited conditions. These results confirm that coordinated mobility and power control can effectively mitigate interference and improve connectivity in highly dynamic vehicular networks, paving the way for robust communication in future autonomous and UAV systems.